\documentclass[conference]{IEEEtran} 
\IEEEoverridecommandlockouts
\pdfoutput=1

\usepackage{amsmath,graphicx,hyperref}
\usepackage{cite}
\usepackage{amsmath,amssymb,amsfonts}
\usepackage{algorithmic}
\usepackage{graphicx}
\usepackage{textcomp}
\usepackage{xcolor}
\usepackage{bbm}
\usepackage{siunitx}
\usepackage{pgf, tikz, pgfplots}
\usepgfplotslibrary{groupplots}
\usetikzlibrary{shapes, arrows.meta, calc, matrix, patterns, positioning}
\usetikzlibrary{spy,decorations.pathreplacing,calligraphy,calc,arrows, arrows.meta, dsp,matrix,patterns}
\usepackage{pgfplots}
\usepgfplotslibrary{groupplots,fillbetween}
\tikzset{>=latex}
\usepackage[ruled]{algorithm2e}
\usepackage{todonotes}
\usepackage{mleftright}
\usepackage[nolist]{acronym}
\renewcommand{\vec}[1]{\mathbf{#1}}
\newcommand{\vecs}[1]{\boldsymbol{#1}}

\newcommand{\hv}{\vec{h}}

\newcommand{\mv}{\vec{m}}
\newcommand{\nv}{\vec{n}}

\newcommand{\vv}{\vec{v}}

\newcommand{\xv}{\vec{x}}
\newcommand{\yv}{\vec{y}}

\newcommand{\gammav}{\vecs{\gamma}}

\newcommand{\muv}{\vecs{\mu}}

\newcommand{\sigmav}{\vecs{\sigma}}

\newcommand{\xFv}{\vec{x}^\vec{F}}

\newcommand{\xIv}{\vec{x}^\vec{I}}

\newcommand{\Fm}{\vec{F}}

\newcommand{\Hm}{\vec{H}}
\newcommand{\Id}{\vec{I}}

\newcommand{\Vm}{\vec{V}}
\newcommand{\Wm}{\vec{W}}

\newcommand{\Sigmam}{\vecs{\Sigma}}

\newcommand{\Nc}{{\cal N}}

\newcommand{\RR}{\mathbb{R}}

\begin{acronym}
 \acro{CSI}{channel state information}
 \acro{NBP}{neural belief propagation}
 \acro{N-SPA}{neural sum-product algorithm}
 \acro{FFG}{Forney factor graph}
 \acro{UFG}{Ungerboeck factor graph}
 \acro{PCM}{parity-check matrix}
 \acro{UE}{user equipment}
 \acro{UL}{uplink}
 \acro{BS}{basestation}
 \acro{TDD}{turbo detection and decoding}
 \acro{TDL}{tapped delay line}
 \acro{FDD}{frequency division duplex}
 \acro{ECC}{error-correcting code}
 \acro{MLD}{maximum likelihood decoding}
 \acro{HDD}{hard decision decoding}
 \acro{IF}{intermediate frequency}
 \acro{RF}{radio frequency}
 \acro{SDD}{soft decision decoding}
 \acro{NND}{neural network decoding}
 \acro{CNN}{convolutional neural network}
 \acro{ML}{maximum likelihood}
 \acro{GPU}{graphical processing unit}
 \acro{BP}{belief propagation}
 \acro{LTE}{Long Term Evolution}
 \acro{BER}{bit error rate}
 \acro{SER}{symbol error rate}
 \acro{DER}{detection error rate}
 \acro{SNR}{signal-to-noise-ratio}
 \acro{ReLU}{rectified linear unit}
 \acro{BPSK}{binary phase shift keying}
 \acro{QPSK}{quadrature phase shift keying}
 \acro{AWGN}{additive white Gaussian noise}
 \acro{MSE}{mean squared error}
 \acro{LLR}{log-likelihood ratio}
 \acro{MAP}{maximum a posteriori}
 \acro{NVE}{normalized validation error}
 \acro{BCE}{binary cross-entropy}
 \acro{CE}{cross-entropy}
 \acro{BLER}{block error rate}
 \acro{SQR}{signal-to-quantisation-noise-ratio}
 \acro{MIMO}{multiple-input multiple-output}
 \acro{OFDM}{orthogonal frequency division multiplex}
 \acro{RF}{radio frequency}
 \acro{LOS}{line of sight}
 \acro{NLoS}{non-line of sight}
 \acro{NMSE}{normalized mean squared error}
 \acro{CFO}{carrier frequency offset}
 \acro{SFO}{sampling frequency offset}
 \acro{IPS}{indoor positioning system}
 \acro{TRIPS}{time-reversal IPS}
 \acro{RSSI}{received signal strength indicator}
 \acro{MIMO}{multiple-input multiple-output}
 \acro{ENoB}{effective number of bits}
 \acro{AGC}{automated gain control}
 \acro{ADC}{analog to digital converter}
 \acro{ADCs}{analog to digital converters}
 \acro{FB}{front bandpass}
 \acro{FPGA}{field programmable gate array}
 \acro{JSDM}{Joint Spatial Division and Multiplexing}
 \acro{NN}{neural network}
 \acro{IF}{intermediate frequency}
 \acro{LoS}{line-of-sight}
 \acro{NLoS}{non-line-of-sight}
 \acro{DSP}{digital signal processing}
 \acro{AFE}{analog front end}
 \acro{SQNR}{signal-to-quantisation-noise-ratio}
 \acro{SINR}{signal-to-interference-noise-ratio}
 \acro{ENoB}{effective number of bits}
 \acro{AGC}{automated gain control}
 \acro{PCB}{printed circuit board}
 \acro{EVM}{error vector mangnitude}
 \acro{CDF}{cumulative distribution function}
 \acro{MRC}{maximum ratio combining}
 \acro{MRP}{maximum ratio precoding}
 \acro{MRT}{maximum ratio transmission}
 \acro{DeepL}{deep-learning}
 \acro{DL}{deep learning}
 \acro{SISO}{single-input single-output}
 \acro{SGD}{stochastic gradient descent}
 \acro{CP}{cyclic prefix}
 \acro{MISO}{Multiple Input Single Output}
 \acro{LMMSE}{linear minimum mean square error}
 \acro{ZF}{zero forcing}
 \acro{USRP}{universal software radio peripheral}
 \acro{RNN}{recurrent neural network}
 \acro{GRU}{gated recurrent unit}
 \acro{LSTM}{long short-term memory}
 \acro{NTM}{neural turing machine}
 \acro{DNC}{differentiable neural computer}
 \acro{TCN}{temporal convolutional network}
 \acro{FCL}{fully connected layer}
 \acro{MLP}{multilayer perceptron}
 \acro{MANN}{memory augmented neural network}
 \acro{RNN}{recurrent neural network}
 \acro{DNN}{deep neural network}
 \acro{FIR}{finite impulse response}
 \acro{BPTT}{back-propagation through time}
 \acro{GAN}{generative adversarial network}
 \acro{ELU}{exponential linear unit}
 \acro{tanh}{hyperbolic tangent}
 \acro{BICM}{bit-interleaved coded modulation}
 \acro{OTA}{over-the-air}
 \acro{IM}{intensity modulation}
 \acro{DD}{direct detection}
 \acro{RL}{reinforcement learning}
 \acro{SDR}{software-defined radio}
 \acro{WGAN}{Wasserstein generative adversarial network}
 \acro{BMD}{bit-metric decoding}
 \acro{BMI}{bit-wise mutual information}
 \acro{SMI}{symbol-wise mutual information}
 \acro{LDPC}{low-density parity-check}
 \acro{IDD}{iterative demapping and decoding}
 \acro{IEDD}{iterative equalization, demapping and decoding}
 \acro{JDD}{joint detection and decoding}
 \acro{SDD}{separate detection and decoding}
 \acro{JSD}{Jensen-Shannon divergence}
  \acro{FLOP}{floating point operation}
 \acro{MMSE}{minimum mean square error}
 \acro{FFT}{fast Fourier transform}
 \acro{IFFT}{inverse fast Fourier transform}
 \acro{QAM}{quadrature amplitude modulation}
 \acro{PAM}{pulse amplitude modulation}
 \acro{EMD}{earth mover's distance}
 \acro{TDL}{tapped delay line}
 \acro{KL}{Kullback--Leibler}
 \acro{PRACH}{physical random access channel}
 \acro{URLLC}{ultra-reliable low-latency communication}
 \acro{ANOMA}{asynchronous non-orthogonal multiple access}
 \acro{FEC}{forward error correction}
 \acro{NOMA}{non-orthogonal medium access}
 \acro{MTC}{machine-type communications}
 \acro{mMTC}{massive machine-type communications}
 \acro{MCS}{modulation and coding scheme}
 \acro{PAPR}{peak-to-average power ratio}
 \acro{MAC}{medium access control}
 \acro{STO}{sampling time offset}
 \acro{STE}{straight-through estimator}
 \acro{PHY}{physical}
 \acro{CCE}{categorical cross-entropy}
 \acro{IoT}{Internet of Things}
 \acro{CCDF}{complementary cumulative distribution function}
 \acro{CRC}{cyclic redundancy check}
 \acro{ACLR}{adjacent channel leakage ratio}
\acro{MD}{missed detection}
 \acro{FA}{false alarm}
 \acro{FAR}{false alarm rate}
 \acro{BCJR}{Bahl--Cocke--Jelinek--Raviv}
 \acro{SCNC}{serially concatenated neural code}
 \acro{GNN}{graph neural network}
 \acro{VN}{variable node}
 \acro{CN}{check node}
 \acro{FN}{factor node}
 \acro{DUIDD}{deep-unfolded interleaved detection and decoding}
 \acro{ISI}{inter-symbol interference}
\acro{APP}{a posteriori probability}
\acro{EXIT}{extrinsic information transfer}
\acro{EP}{expectation propagation}
\acro{CS}{channel shortening}
\acro{FGNN}{factor graph neural network}
\acro{FCN}{fully connected network}
\acro{DCCB-SC}{deep
concatenated convolutional blocks with skip connections}
\acro{SPA}{sum-product algorithm}
\acro{CIR}{channel impulse response}
\acro{MAC}{multiply-accumulate}
\acro{CCT}{constant channel transformation}
\acro{GAT}{graph attention network}
\acro{LE}{linear estimator}
\acro{NLE}{non-linear estimator}
\acro{PDF}{probability density function}
\acro{PMF}{probability mass function}
\acro{KLD}{Kullback-Leibler divergence}
\acro{i.i.d.}{independent and identically
distributed}
\acro{DICHASUS}{Channel Sounder by University of Stuttgart}
\end{acronym}

\definecolor{mittelblau}{RGB}{0, 126, 198}
\definecolor{violettblau}{cmyk}{0.9, 0.6, 0, 0}
\definecolor{rot}{RGB}{238, 28 35}
\definecolor{apfelgruen}{RGB}{140, 198, 62}
\definecolor{gelb}{RGB}{1, 221, 0}
\definecolor{orange}{RGB}{244, 111, 33}
\definecolor{pink}{RGB}{237, 0, 140}
\definecolor{lila}{RGB}{128, 10, 145}
\definecolor{hellgrau}{RGB}{224, 224, 224}
\definecolor{mittelgrau}{RGB}{128, 128, 128}
\definecolor{dunkelgrau}{RGB}{80,80,80}
\definecolor{anthrazit}{RGB}{19, 31, 31}

\pgfplotscreateplotcyclelist{corporate colours markers}{%
rot, every mark/.append style={fill=.!80!rot},mark=*\\%
mittelblau, every mark/.append style={fill=.!80!mittelblau},mark=square*\\%
apfelgruen, every mark/.append style={fill=.!80!apfelgruen},mark=triangle*\\%
orange, mark=star\\%
pink, every mark/.append style={fill=.!80!pink},mark=diamond*\\%
violettblau, every mark/.append style={fill=.!80!violettblau},mark=otimes*\\%
lila, mark=|\\%
gelb, every mark/.append style={fill=.!80!gelb},mark=pentagon*\\%
hellgrau, mark=text,text mark=p\\%
anthrazit, mark=text,text mark=a\\%
}

\pgfplotscreateplotcyclelist{corporate colours markers double}{%
rot, every mark/.append style={fill=.!80!rot},mark=*\\%
rot, dashed, every mark/.append style={fill=.!80!rot,solid},mark=*\\%
mittelblau, every mark/.append style={fill=.!80!mittelblau},mark=square*\\%
mittelblau, dashed, every mark/.append style={fill=.!80!mittelblau,solid},mark=square*\\%
apfelgruen, every mark/.append style={fill=.!80!apfelgruen},mark=triangle*\\%
apfelgruen, dashed, every mark/.append style={fill=.!80!apfelgruen,solid},mark=triangle*\\%
orange, mark=star\\%
orange, dashed, mark=star\\%
pink, every mark/.append style={fill=.!80!pink},mark=diamond*\\%
pink, dashed, every mark/.append style={fill=.!80!pink,solid},mark=diamond*\\%
violettblau, every mark/.append style={fill=.!80!violettblau},mark=otimes*\\%
violettblau, dashed, every mark/.append style={fill=.!80!violettblau,solid},mark=otimes*\\%
lila, mark=|\\%
lila, dashed, mark=|\\%
gelb, every mark/.append style={fill=.!80!gelb},mark=pentagon*\\%
gelb, dashed, every mark/.append style={fill=.!80!gelb,solid},mark=pentagon*\\%
}

\pgfplotsset{
  colormap/magma/.style={%
    /pgfplots/colormap={magma}{%
      rgb=(0.001462, 0.000466, 0.013866)
      rgb=(0.035520, 0.028397, 0.125209)
      rgb=(0.102815, 0.063010, 0.257854)
      rgb=(0.191460, 0.064818, 0.396152)
      rgb=(0.291366, 0.064553, 0.475462)
      rgb=(0.384299, 0.097855, 0.501002)
      rgb=(0.475780, 0.134577, 0.507921)
      rgb=(0.569172, 0.167454, 0.504105)
      rgb=(0.664915, 0.198075, 0.488836)
      rgb=(0.761077, 0.231214, 0.460162)
      rgb=(0.852126, 0.276106, 0.418573)
      rgb=(0.925937, 0.346844, 0.374959)
      rgb=(0.969680, 0.446936, 0.360311)
      rgb=(0.989363, 0.557873, 0.391671)
      rgb=(0.996580, 0.668256, 0.456192)
      rgb=(0.996727, 0.776795, 0.541039)
      rgb=(0.992440, 0.884330, 0.640099)
      rgb=(0.987053, 0.991438, 0.749504)
    },
  },
  colormap/inferno/.style={%
    /pgfplots/colormap={inferno}{%
      rgb=(0.001462, 0.000466, 0.013866)
      rgb=(0.037668, 0.025921, 0.132232)
      rgb=(0.116656, 0.047574, 0.272321)
      rgb=(0.217949, 0.036615, 0.383522)
      rgb=(0.316282, 0.053490, 0.425116)
      rgb=(0.410113, 0.087896, 0.433098)
      rgb=(0.503493, 0.121575, 0.423356)
      rgb=(0.596940, 0.154848, 0.398125)
      rgb=(0.688653, 0.192239, 0.357603)
      rgb=(0.775059, 0.239667, 0.303526)
      rgb=(0.851384, 0.302260, 0.239636)
      rgb=(0.912966, 0.381636, 0.169755)
      rgb=(0.956852, 0.475356, 0.094695)
      rgb=(0.981895, 0.579392, 0.026250)
      rgb=(0.987464, 0.690366, 0.079990)
      rgb=(0.973088, 0.805409, 0.216877)
      rgb=(0.947594, 0.917399, 0.410665)
      rgb=(0.988362, 0.998364, 0.644924)
    },
  },
  colormap/plasma/.style={%
    /pgfplots/colormap={plasma}{%
      rgb=(0.050383, 0.029803, 0.527975)
      rgb=(0.186213, 0.018803, 0.587228)
      rgb=(0.287076, 0.010855, 0.627295)
      rgb=(0.381047, 0.001814, 0.653068)
      rgb=(0.471457, 0.005678, 0.659897)
      rgb=(0.557243, 0.047331, 0.643443)
      rgb=(0.636008, 0.112092, 0.605205)
      rgb=(0.706178, 0.178437, 0.553657)
      rgb=(0.768090, 0.244817, 0.498465)
      rgb=(0.823132, 0.311261, 0.444806)
      rgb=(0.872303, 0.378774, 0.393355)
      rgb=(0.915471, 0.448807, 0.342890)
      rgb=(0.951344, 0.522850, 0.292275)
      rgb=(0.977856, 0.602051, 0.241387)
      rgb=(0.992541, 0.687030, 0.192170)
      rgb=(0.992505, 0.777967, 0.152855)
      rgb=(0.974443, 0.874622, 0.144061)
      rgb=(0.940015, 0.975158, 0.131326)
    },
  },
}

\def\BibTeX{{\rm B\kern-.05em{\sc i\kern-.025em b}\kern-.08em
    T\kern-.1667em\lower.7ex\hbox{E}\kern-.125emX}}
\begin{document}

\newpage

\title{Signal  Space-Transformed Expectation Propagation
for Symbol Detection  in ISI Channels}

\author{\IEEEauthorblockN{Jannis Clausius\IEEEauthorrefmark{1}  \qquad Luca Schmid\IEEEauthorrefmark{2} \qquad Laurent Schmalen\IEEEauthorrefmark{2} \qquad Stephan ten Brink\IEEEauthorrefmark{1}} \thanks{This work is supported in part from the German Federal Ministry of Research, Technology and Space (BMFTR) within the project Open6GHub (16KISK019, 16KISK010), Open6GHubPlus/FKZ (16KIS2406, 16KIS2405) and in part from the European Research Council (ERC) under the European Union’s Horizon 2020 research and innovation programme (101001899).}

	\IEEEauthorblockA{
		\IEEEauthorrefmark{1} Institute of Telecommunications, Pfaffenwaldring 47, University of  Stuttgart, 70569 Stuttgart, Germany \\
        \IEEEauthorrefmark{2} Communications Engineering Lab, Karlsruhe Institute of Technology, 76187 Karlsruhe, Germany
		\\Email: clausius@inue.uni-stuttgart.de\\
	}
}

\maketitle

\begin{abstract}
Iterative message passing detection based on \ac{EP} has demonstrated near-optimum performance in many signal processing and communication scenarios. The method remains feasible even for \acp{CIR}, where the optimal \ac{BCJR} detector is infeasible.
However, significant performance degradation occurs for channels with strong \ac{ISI}, where the initial \ac{LMMSE} estimate is inaccurate.
We propose an \ac{EP}-based detector that operates in a transformed signal space.
Specifically, instead of the conventional approach that iterates between an \ac{LMMSE} estimator and a non-linear symbol-wise demapper, the proposed method iterates between a linear channel shortening filter-based estimator and a non-linear \ac{BCJR} detector with reduced memory compared to the actual channel.
Additionally, we propose a deliberate mismatch between the initialized messages and the initialized covariance used in the linear estimator in the first iteration for faster convergence.
The proposed approach is evaluated for the well-known Proakis-C \ac{ISI} channel and for \acp{CIR} from a wireless measurement campaign.  We demonstrate improvements of up to 6\,dB at  2\,bits per channel use and an improved performance-complexity trade-off over conventional EP-based detection.

\end{abstract}
\acresetall
%

\section{Introduction}

\Acf{EP} is a general framework for approximate Bayesian inference~\cite{minka_expectation_2013}.
The product of the likelihood and the prior is approximated with an exponential family distribution, usually a Gaussian, and iteratively refined.
The adaptation of \ac{EP} to the communication scenario over \ac{MIMO} and \ac{ISI} channels \cite{ep_senst,ep_cespedes,santos2015block} can be interpreted as an iterative message passing algorithm between a \ac{LMMSE} estimator and a symbol-wise demapper.
More generally, messages are exchanged between a \ac{LE} and a \ac{NLE}.
In many cases, the true posterior can be closely approximated and near-optimal performance is achieved.
Moreover, \ac{EP}-based detection remains feasible for long \acp{CIR} where the optimal \ac{BCJR} detector \cite{bahl1974optimal} is infeasible due to computational complexity.
However, in the case of discrete priors and a poor initial estimate of the \ac{LE}, \ac{EP} may fail to well-approximate the true posterior. 
Consequently, improvements were proposed, e.g., carefully fine-tuning the hyperparameters \cite{turboep17santos,santos_turbo_2018,improve_ep_yao_21},  adding neural components \cite{Kosasih22GnnMIMO}, or passing messages based on a Gaussian mixture model \cite{shayovitz2024mimo}.


Here, we propose the application of \ac{EP} in a transformed signal space obtained by linear channel shortening \cite{falconer_adaptive_1973}.
The transformation to this signal space enhances the initialization of the \ac{LE} and the \ac{NLE}, which now operate in the transformed space.
The resulting transformed system model introduces memory, albeit less than the original channel.
To account for the memory, the symbol-wise demapper is replaced by a more suitable \ac{NLE}, i.e., a trellis-based \ac{BCJR} detector.
Note that the \ac{BCJR} detector is based on the reduced memory of the transformed system model and does not scale exponentially in complexity with the memory of the original channel.
The memory of the transformed system model and the number of message passing iterations are hyperparameters that enable a fine-granular trade-off between performance and complexity.

\section{Preliminaries}
\vspace{-0.25em}
\subsection{System Model}
We consider a linear system model
\vspace{-0.25em}
\begin{equation}\label{eq:system}
    \tilde{\yv} = \tilde{\Hm} \tilde{\xv }+ \tilde{\nv},
    \qquad \tilde{\nv}\sim \mathcal{C}\Nc(\boldsymbol{0}, \tilde{N}_0\Id) \nonumber
\end{equation}
\vspace{-0.25em}
with uniform, \ac{i.i.d.} transmit symbols $\tilde{\xv } \in \tilde{\mathcal{X}}^N \subset  \mathbb{C}^N$ from an alphabet $\tilde{\mathcal{X}}$ with $|\tilde{\mathcal{X}}| = M$.
For \ac{ISI} channels, the normalized channel matrix $\tilde{\Hm} \in \mathbb{C}^{(N+L)\times N}$ is a lower triangular Toeplitz matrix with memory $L\geq0$, i.e., ${H_{i,j} = 0}$ for ${L<i-j}$.
This is equivalent to a convolution with a static \ac{CIR}~$\tilde{\hv}$ of length $L+1$.
The equivalent real-valued system model is
\vspace{-0.25em}
\begin{equation}
    \yv = \Hm \xv +\nv
    \label{eq:yHxn}
\end{equation}
\vspace{-0.25em}
with the decomposed components  
\vspace{-0.25em}
\begin{equation}\label{eq:decompose}
    \Hm = 
    \begin{bmatrix}
        \mathcal{R}(\tilde{\Hm}) & -\mathcal{I}(\tilde{\Hm}) \\
        \mathcal{I}(\tilde{\Hm}) & \mathcal{R}(\tilde{\Hm})
    \end{bmatrix}, \;
    \xv = [\mathcal{R}(\xv)^\mathrm{T}, \mathcal{I}(\xv)^\mathrm{T}]^\mathrm{T}
\end{equation}
\vspace{-0.25em}
and real-valued noise~$\nv\sim \Nc(\boldsymbol{0}, N_0\Id)$
with $N_0 = \tilde{N}_0 / 2$  and \ac{SNR} $E_\mathrm{S}/N_0$.
The task of symbol detection at the receiver can be approached via \ac{MAP} detection
\vspace{-0.25em}
\begin{equation}
    \hat{x}_i = \arg\max_{x_i} P(x_i|\yv) \propto \arg\max_{x_i} \sum_{\sim \{x_i\}} p(\yv|\xv)\cdot P(\xv) \label{eq:map_detection}
\end{equation}
\vspace{-0.25em}
where ${\sim \! \{x_i\}}$ indicates the set of all $x_j$ with ${j\neq i}$.
We are interested in the distribution $P(\hat{x}_i)$ that corresponds to the soft output of a detector.
The \ac{SMI} ${I_\mathrm{SMI}(\hat{\xv}^{\Id};\xIv)=1/(2N)\sum_iI(x_i,\hat{x}_i)}$ evaluates the quality of the distribution in terms of the information rate.

\vspace{-0.25em}
\subsection{BCJR Detector}\label{sec:BCJR}
The \ac{BCJR} algorithm \cite{bahl1974optimal} is an optimal detector for the system model (\ref{eq:yHxn}). 
It relies on the forward and backward recursions~\cite{dlr6855} through a trellis.
The branches of the trellis are initialized by the log-likelihoods $\gamma_i \, \forall i=0,1,...,N+L-1$ and updated during recursion to $\gamma'_i$.
Through a final marginalization of $\gamma'_i$, the algorithm produces symbol-wise \ac{APP} estimates.
The complexity scales with $\mathcal{O}(M^{L+1})$ and thus is only feasible for a small memory~$L$.

\vspace{-0.25em}
\subsection{Channel Shortening} \label{sec:shortening}
A channel shortening filter \cite{falconer_adaptive_1973,rusek12optimalshortening} is a linear receive filter $\Wm$ that transforms the system model~\eqref{eq:yHxn} into 
\vspace{-0.25em}
\begin{equation}\label{eq:cs}
     \yv^\Fm=\Wm\Hm\xv+\boldsymbol{\epsilon}_\Fm, \quad \text{with} \quad \Fm\approx \Wm\Hm, 
\end{equation}
\vspace{-0.25em}
where $\Fm$ is a target channel with constrained memory, i.e., $\nu < L$ and ${F_{i,j} = 0}$ for ${\nu<i-j}$. The observed variable in the signal space defined by $\Fm$ is denoted with $\yv^\Fm$ and $\boldsymbol{\epsilon}^\Fm$ is an error term.
In the case of \ac{ISI} channels, $\Fm$ is Toeplitz and lower triangular. 
The optimal filter, in the \ac{MSE} sense, is
\vspace{-0.25em}
\begin{equation}
    \Wm = \Fm \left( \frac{1}{N_0}  \Hm^\mathrm{T} \Hm + \Id \right)^{-1} \frac{1}{N_0} \Hm^\mathrm{T}. \nonumber
\end{equation}
\vspace{-0.25em}
Subsequently, the new, transformed system model in (\ref{eq:cs}) with reduced memory $\nu < L$ can be processed by an \ac{NLE} with reduced complexity, i.e., a \ac{BCJR} detector with complexity $\mathcal{O}(M^{\nu+1})$.
If the target channel is memoryless ($\Fm=\Id$), the \ac{LMMSE} filter follows \cite{proakis2001digital}.

\vspace{-0.25em}
\subsection{Expectation Propagation}\label{sec:EP}

\Ac{EP} is a general framework for approximate Bayesian inference~\cite{minka_expectation_2013}. For our application, it is convenient to view \ac{EP} as a message passing scheme iterating between an \ac{LE} and an \ac{NLE} \cite{ep_cespedes,ep_senst, Fischer2022}.
The \ac{LE} and the \ac{NLE} are derived by assuming that the prior and the likelihood in \eqref{eq:map_detection} are fully-factorized Gaussians $t_\mathrm{A}(\xv) = \prod_i t_i(x_i)\sim \Nc(\xv: \mv_\mathrm{A}, \operatorname{diag}(\vv_\mathrm{A})) $ and $q_\mathrm{A}(\xv)= \prod_i q_i(x_i)\sim\Nc(\xv: \muv_\mathrm{A}, \operatorname{diag}(\sigmav_\mathrm{A}^2)) $, respectively. The subscript $\mathrm{A}$ indicates \emph{a priori} knowledge available from the other component.
The resulting estimators are
\vspace{-0.25em}
\begin{align}
    \hat{p}_\mathrm{LE}(\xv)&=p(\yv|\xv)\cdot t_{\mathrm{A}}(\xv), \nonumber \\
    \hat{P}_\mathrm{NLE}(\xv)&= q_{\mathrm{A}}(\xv) \cdot P(\xv). \nonumber
\end{align}
\vspace{-0.25em}
Subsequently, each estimator projects its respective result onto a fully-factorized Gaussian
\begin{equation}
    \operatorname{Proj}_i(p(\xv)) = \Nc\Big(\operatorname{E}\Big(\sum_{\sim \{x_i\}}p(\xv)\Big) ,\operatorname{Var}\Big(\sum_{\sim \{x_i\}}p(\xv)\Big)\Big), \; \forall i \nonumber
\end{equation}
\vspace{-0.25em}
where $\operatorname{E}(\cdot)$ is the expectation, and $\operatorname{Var}(\cdot)$ is the variance.
The projection consists of two steps. 
First, the marginalized \ac{PDF}/\ac{PMF} $\sum_{\sim \{x_i\}}p(\xv)$ is calculated.
Second, the \ac{KLD} between the marginalized \ac{PDF}/\ac{PMF} and a fully-factorized Gaussian is minimized using moment matching~\cite{minka_expectation_2013}.
As a result, only the moments of the messages $q_i(x_i)=\operatorname{Proj}_i(\hat{p}_\mathrm{LE}(\xv))\sim\Nc(x_i: \mu_i, \sigma_i)$ and $t_i(x_i)=\operatorname{Proj}_i(\hat{P}_\mathrm{NLE}(\xv))\sim\Nc(x_i: m_i, v_i)$ are passed between the \ac{NLE} and \ac{LE}.
The Gaussian projections are divided by the a priori knowledge of the other estimator to obtain extrinsics
\vspace{-.3em}
\begin{equation}
    q_{\mathrm{A},i}(x_i) = \frac{q_{i}(x_i)}{t_{\mathrm{A},i}(x_i)}, \qquad t_{\mathrm{A},i}(x_i) = \frac{t_{i}(x_i)}{q_{\mathrm{A},i}(x_i)}. \nonumber
\end{equation} 
\vspace{-0.25em}
These messages can be seen as a priori knowledge and are also referred to as extrinsic information, cavity distribution, or unbiased estimate.
The corresponding moments are
\vspace{-.3em}
\begin{align}
     \sigma^2_{\mathrm{A},i} &=\left(\frac{1}{\sigma^2_{i}} - \frac{1}{v_{\mathrm{A},i}}   \right)^{-1}, \qquad \mu_{\mathrm{A},i} =\sigma^2_{\mathrm{A},i} \left(\frac{\mu_i}{\sigma^2_{i}} - \frac{m_{\mathrm{A},i}}{v_{\mathrm{A},i}}   \right), \nonumber \\
            v'_{\mathrm{A},i} &=\bigg(\frac{1}{v_{i}} - \frac{1}{\sigma^2_{\mathrm{A},i}}   \bigg)^{-1},  \qquad
            m'_{\mathrm{A},i} =v'_{\mathrm{A},i} \bigg(\frac{m_{i}}{v_{i}} - \frac{\mu_{\mathrm{A},i}}{\sigma^2_{\mathrm{A},i}} \bigg).\nonumber
\end{align}
\vspace{-0.25em}
Finally, convergence is aided by momentum-based updates
\vspace{-.3em}
\begin{align}
    v^{(\ell)}_{\mathrm{A},i} =&\left(\frac{\beta  }{v'_{\mathrm{A},i}} + \frac{1-\beta  }{v^{(\ell-1)}_{\mathrm{A},i} } \right), \nonumber
    \\
    m^{(\ell)}_{\mathrm{A},i} =&\left(\beta  m'_{\mathrm{A},i} + (1-\beta  ) m^{(\ell-1)}_{\mathrm{A},i}  \right),\nonumber
\end{align}
\vspace{-0.25em}
where ${\beta\in (0,1)}$ is the step size, $\ell\in \left[0,N_\mathrm{It}-1 \right]$ indicates the iteration, and $N_\mathrm{It}$ is the number of iterations.

\section{Expectation Propagation in a Transformed Signal Space}

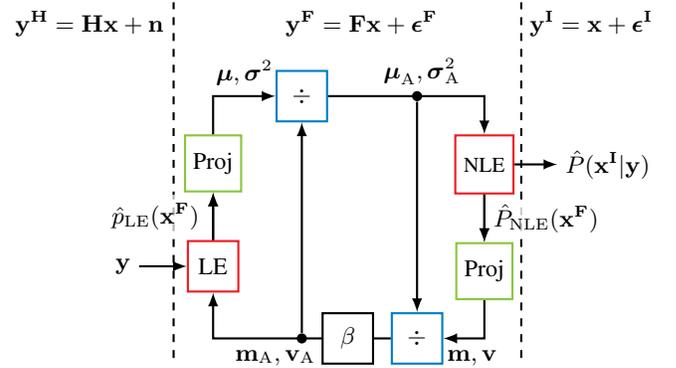
\begin{figure}[t]
    \centering
    \tikzstyle{block} = [draw,  thick, regular polygon, regular polygon sides=4, minimum width = 3em, inner sep=0pt]
    \tikzstyle{rblock} = [draw,  thick, rectangle, minimum width=3em,  minimum height=1.5em, inner sep=4pt]
    \tikzstyle{arrow} = [-latex, thick]
    \begin{tikzpicture}
        \node[block, draw=rot](F){\normalsize LE};
        \node[left =2em of F](F_in){\footnotesize $\yv$};
        \node[block, above =1.5em of F, draw=apfelgruen](marg){\normalsize Proj};
        
        \node[block, above right=0.1em and 1.0em of marg, draw=mittelblau](cav1){$\div$};
        
        \node[block, below right=0.25em and 5.0em of cav1, align=left, draw=rot](det){\footnotesize NLE};
        \node[right =1.8em of det](det_out){\footnotesize $\hat{P}(\xIv|\yv)$};
        \node[block, below=1.5em of det , draw=apfelgruen](proj){\normalsize Proj};
        
        \node[block, below left=0.1em and 0.5em of proj, draw=mittelblau](cav2){$\div$};
        
        \node[fill,circle, anchor=center, inner sep=0.15em] at (cav1-|cav2) (c1){};
        \node[fill,circle, anchor=center, inner sep=0.15em] at (cav2-|cav1) (c2){};

        \draw[arrow] (F_in) -- (F);
        \draw[arrow] (F) -- (marg);
        \draw[arrow] (marg) |-node[above=0.1em, pos=0.75]{\footnotesize $ \muv, \sigmav^2$} (cav1);
        \draw[arrow] (cav1) -|node[above=0.1em, pos=0.3]{\footnotesize $ \muv_{\mathrm{A}}, \sigmav^2_{\mathrm{A}}$} (det);
        \draw[arrow] (det) -- (proj);
        \draw[arrow] (proj) |- node[below=0.1em, pos=0.65]{\footnotesize $ \mv, \vv$} (cav2);
        \draw[arrow] (cav2) -|node[below=0.1em, pos=0.345]{\footnotesize $\mv_{\mathrm{A}}, \vv_{\mathrm{A}}$} (F);
        \draw[arrow] (c1) -- (cav2);
        \draw[arrow] (c2) -- (cav1);
         \draw[arrow ] (det) -- (det_out);
        \node[block, left=0.7em  of cav2, fill=white](momentum){$\beta$};

        \node[ below left=2.75em and 0.2em of F](l_u){};
        \node[ above left=9em and 0.2em of F](l_o){};

        \node[  right=0em and 13.2em of l_u](r_u){};
        \node[  right=0em and 13.2em of l_o](r_o){};

        \node[ below right=0em and -0.35em of r_o](){ \small $\yv^\mathrm{\Id} = \xv+\boldsymbol{\epsilon}^\Id$};
        
        \node[ below left=0em and -0.3em of l_o](){\small $\yv^\Hm = \Hm \xv+\nv$};

        \node[ below right=0em and 4.em of l_o](){\small $\yv^\mathrm{\Fm} = \Fm \xv+\boldsymbol{\epsilon}^\Fm$};

        \draw[dashed, thick] (l_u) -- (l_o);
        \draw[dashed, thick] (r_u) -- (r_o);

        \draw[arrow] (F) -- node[left=0.2em, pos=0.5,rectangle, fill=white, fill opacity=0.7, text opacity=1.0]{\footnotesize $\hat{p}_\mathrm{LE}(\xFv)$} (marg);
        \draw[arrow] (det) -- node[right=0.2em, pos=0.5,rectangle,inner sep=0.15em, fill=white, fill opacity=0.8,text opacity=1.0]{\footnotesize $\hat{P}_\mathrm{NLE}(\xFv)$}(proj);

    \end{tikzpicture}
    \vspace{-2.5em}
    \caption{\small Block diagram of the proposed scheme. The \ac{EP} framework iterates between a \acf{LE} and a \acf{NLE}, but in a transformed signal space defined by $\Fm$; the system model is indicated at the top. Estimators are colored in red, the projections onto fully-factorized Gaussians are green, the \ac{PDF} divisions are blue, and the momentum is black. 
     }
    \label{fig:bcjrep_block_diagram}
    \vspace{-1.0em}
\end{figure}


As shown in Fig.~\ref{fig:bcjrep_block_diagram}, we propose to perform \ac{EP} from the perspective of the transformed signal space $\xFv = \Fm \xIv \in \mathcal{X}^\Fm$, where $\Fm$ is a linear transformation, $\xIv=\xv$ are the transmitted symbols, and $\mathcal{X}^\Fm$ is the transformed alphabet with cardinality $|\mathcal{X}^\Fm|=M^{\nu+1}$. The superscript indicates the signal space of $\xv$.
The conventional \ac{EP} detector is given for ${\Fm = \Id}$. However, we propose to apply the \ac{EP} detector in a transformed signal space $\Fm \neq \Id$. 
This means that the \ac{LE} transforms the system model $\yv^\Hm=\Hm\xIv+\nv$
into $\yv^\Fm = \Fm \xIv + \boldsymbol{\epsilon}^\Fm$ instead of $\yv^\Id =\xIv + \boldsymbol{\epsilon}^\Id$, where $\boldsymbol{\epsilon}^\Fm$ and $\boldsymbol{\epsilon}^\Id$ are error terms. 
Note that the message passing framework still works as described in Sec.~\ref{sec:EP}.
\vspace{-0.25em}
\subsection{Choice of the Transformed Signal Space}\label{sec:choice_F}
The optimal transformed signal space in terms of \ac{SMI} 
is defined by ${\Fm_\mathrm{opt}=\operatorname{arg}\max_\Fm I_\mathrm{SMI}(\hat{\xv}^{\Id};\xIv)}$ where $\hat{\xv}^{\Id}$ is the soft output after detection.
However, the \ac{NLE} must be feasible, which in turn imposes structural constraints on the desired complex channel matrix $\tilde{\Fm}\in\mathbb{C}^{N+\nu \times N}$.
The  real-valued equivalent $\Fm \in \RR^{2(N+\nu)\times 2N}$ can be constructed by decomposition as in \eqref{eq:decompose}. 
We constrain $\tilde{\Fm}$ to be a lower triangular structure with memory $\nu < L$, i.e., ${\tilde{F}_{i,j} = 0}$ for ${\nu<i-j}$, i.e., a shortened channel as discussed in Sec.~\ref{sec:shortening}.
Thus, the optimal \ac{NLE} is a \ac{BCJR} detector with complexity $\mathcal{O}(M^{\nu+1})$.
This enables a spectrum of choices for $\tilde{\Fm}$ with a performance-complexity trade-off ranging from the original \ac{EP} detector~\cite{ep_senst,ep_cespedes} (${\tilde{\Fm} = \Id}$ with ${\nu=0}$) to the optimal \ac{BCJR} detector (${\tilde{\Fm} = \tilde{\Hm}}$) with full memory ${\nu=L}$. 
For varying channel conditions, brute-force optimization of $\Fm_\mathrm{opt}$ is infeasible. 
Instead, we adapt the approach from \cite{rusek12optimalshortening}, which provides a  construction based on a lower bound on the mutual information $I_\mathrm{LB}(\xIv;\yv^\Fm)$.

If one chooses a particular unitary transformation $\Fm$ that defines the transformed signal space and a particular modulation scheme, then the cross domain detector in \cite{Li22cross_domain} follows, where the \ac{LE} and \ac{NLE} are separated by a unitary transformation.
Here, the transformation is part of the \ac{LE} and the \ac{NLE}.

\vspace{-0.25em}
\subsection{Non-Linear Estimator}
The  \ac{NLE} is an \ac{APP} detector operating on the transformed system model $ \yv^\Fm = \Fm \xIv + \boldsymbol{\epsilon}^\Fm$. 
However, to apply the trellis-based \ac{BCJR} detector, the complex-valued transformed signal space induced by the structurally constrained $\tilde{\Fm}\in\mathbb{C}^{N+\nu \times N}$ must be considered. 
Thus, we choose the log-domain branch metrics of the trellis by evaluating $q_i(x_i)\sim\Nc(\xv: \mu_{\mathrm{A},i}, \operatorname{diag}(\sigma_{\mathrm{A},i}^2))$ at all $x^{\tilde{\Fm}}\in\mathcal{X}^{\tilde{\Fm}}$
\vspace{-.5em}
\begin{equation}
    \gamma_i = \frac{\left(\mu_{\mathrm{A},i}-\mathcal{R}(x^{\tilde{\Fm}})\right)^2}{\sigma^2_{\mathrm{A},i}} +\frac{\left(\mu_{\mathrm{A},i+N}-\mathcal{I}(x^{\tilde{\Fm}})\right)^2}{\sigma^2_{\mathrm{A},i+N}}. \nonumber
\end{equation}
\vspace{-0.25em}
The estimate in the transformed signal space is the \ac{PMF} $\hat{P}_\mathrm{NLE}(x^{\tilde{\Fm}}_i)=\operatorname{softmax}(\gammav'_i)$ where $\gammav'_i$ is the branch metric after the forward and backward recursion for all $M^{\nu+1}$ symbols.
Adequately marginalizing $\hat{P}_\mathrm{NLE}(\xv^{\tilde{\Fm}})$ yields $\hat{P}_\mathrm{NLE}(\xv^{\Fm})$.

\begin{figure*}[t]
	\centering
	\input{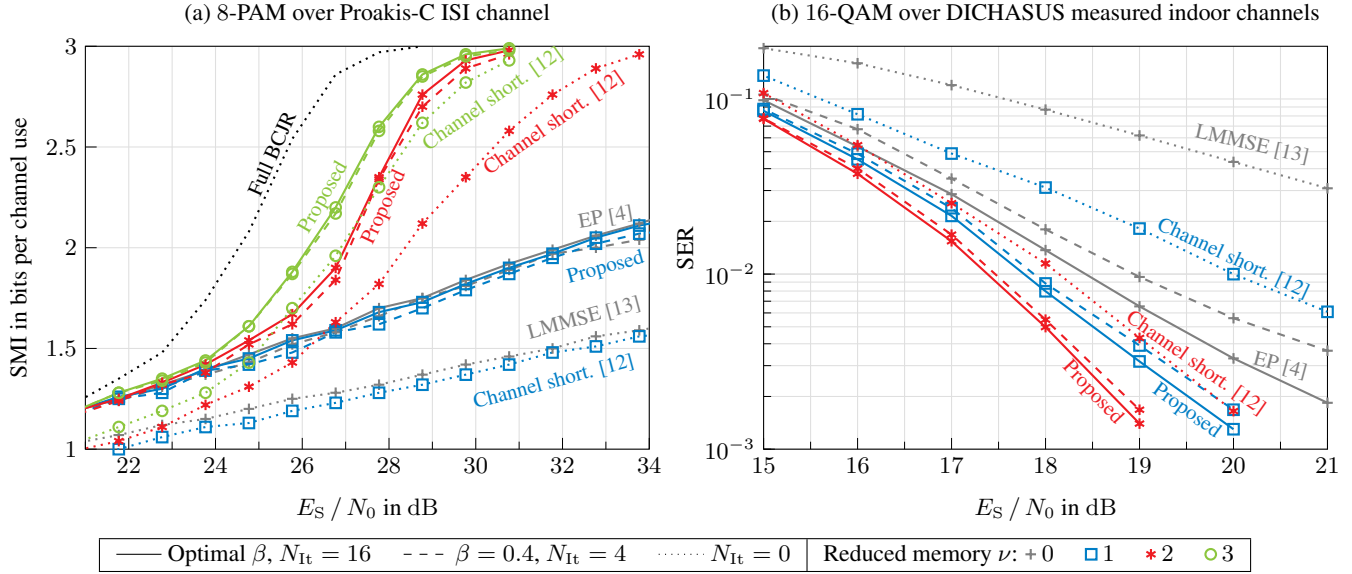}
	\vspace{-0.5em}
	\caption{\small Performance in \acf{SMI} (a) and SER (b) vs. SNR. If memory $\nu>0$ and iterations $N_\mathrm{It}>0$, then the detector is based on EP  in a transformed signal space.
    For $\nu=0$ or $N_\mathrm{It}=0$ the detector is a baseline.
    }
	\label{fig:perfromance}
	\vspace{-1.5em}
\end{figure*}

\vspace{-0.25em}
\subsection{Linear Estimator}\label{sec:LE}

The \ac{LE} operates in the transformed signal space $\xFv$ instead of $\xIv$.
This means we calculate ${p(\yv|\xFv)  t_\mathrm{A}(\xFv) }$ instead of ${p(\yv|\xIv)  \tilde{t}_\mathrm{A}(\xIv)}$ where $\tilde{t}_\mathrm{A}(\cdot)$ is a fully-factorized Gaussian prior with mean $\mv_\mathrm{A}$ and covariance matrix ${\Vm_\mathrm{A} = \operatorname{diag}(\vv_\mathrm{A})}$. 
By using the relationship ${\xIv = \Fm^{-1} \xFv}$, we obtain the likelihood $p(\yv|\xFv) \sim \Nc(\yv:\Hm \Fm^{-1}\xFv, \sigma^2 \Id)$.
From the receiver's perspective, the cascade $\Hm \Fm^{-1}\xFv$ can be interpreted as the transmission of $\xFv$ over the effective channel $\Hm \Fm^{-1}$.
Note that $\Fm$ is not necessarily a square matrix, and the pseudo-inverse ${\left( \Fm^\mathrm{T} \Fm \right)^{-1} \Fm^\mathrm{T} }$ is used.
The resulting posterior is again a Gaussian distribution $p(\yv|\xFv) \cdot t_\mathrm{A}(\xFv) \sim \Nc(\xFv:\muv,\Sigmam)$, with
\vspace{-.5em}
\begin{align}
    \Sigmam &= \left( \frac{1}{N_0} \left(\Fm^{-1}\right)^\mathrm{T} \Hm^\mathrm{T} \Hm \Fm^{-1} + \Vm_\mathrm{A}^{-1} \right)^{-1}, \label{eq:Sigma}\\
    \muv &= \Sigmam \left( \frac{1}{N_0} \left(\Fm^{-1}\right)^\mathrm{T} \Hm^\mathrm{T}\yv +  \Vm_\mathrm{A}^{-1} \mv_\mathrm{A} \right). \label{eq:mu}
\end{align}
\vspace{-0.5em}
\subsection{Mismatched Initialization}\label{sec:convergence}
Assuming only prior knowledge about the constellation alphabet $\mathcal{X}$, i.e., $\xIv$ \ac{i.i.d.} with zero mean and unit energy, then $p(\xFv)\sim \Nc(\xFv: \boldsymbol{0}, \Fm  \Fm^\mathrm{T})$.
Since the \ac{EP} framework assumes a fully-factorized prior, the natural choice is ${\mv^{(0)}_\mathrm{A}=\boldsymbol{0}}$ and ${\vv^{(0)}_\mathrm{A}=\operatorname{diag}(\Fm  \Fm^\mathrm{T})}$ as initialization.
However, using the full covariance matrix in the \ac{LE} can significantly improve performance and simplifies the estimation in the first iteration, as shown below.
Inserting $\Vm_\mathrm{A}=\Fm  \Fm^\mathrm{T}$ as covariance in (\ref{eq:Sigma}) and (\ref{eq:mu}) and simplifying leads to
\vspace{-.5em}
\begin{align*}
    \Sigmam &= \Fm \left( \frac{1}{N_0}  \Hm^\mathrm{T} \Hm + \Id \right)^{-1} \Fm^\mathrm{T},  \\
    \muv &= \Fm \left( \frac{1}{N_0}  \Hm^\mathrm{T} \Hm + \Id \right)^{-1} \left( \frac{1}{N_0}  \Hm^\mathrm{T}\yv  \right). 
\end{align*}
\vspace{-0.25em}
Thereby, the inverse $\Fm^{-1}$ vanishes from the equation, avoiding the potentially poor approximation of the pseudo-inverse and leading to improved estimates.
As a consequence, the classical channel shortening filter from \cite{falconer_adaptive_1973,rusek12optimalshortening} is recovered.
Still, using the full covariance matrix $\Vm_\mathrm{A}=\Fm \Fm^\mathrm{T}$ contradicts the parallel \ac{EP} framework, which assumes a diagonal structure.
Therefore, we propose setting ${\Vm_\mathrm{A}=\Fm \Fm^\mathrm{T}}$ in iteration ${\ell=0}$, but initializing the covariance matrix of the message passing prior as ${\vv^{(0)}_\mathrm{A}  = \operatorname{diag}(\Fm  \Fm^\mathrm{T}) }$, thereby ignoring the slight mismatch. 
Note that the mismatch is twofold: 
If $\Fm  \Fm^\mathrm{T}$ is diagonally dominant, the mismatch effectively vanishes. If $\Fm  \Fm^\mathrm{T}$ is not diagonally dominant, incorporating non-diagonal elements initially yields substantial improvements, such that the mismatch becomes beneficial.

\section{Numerical Results}

\begin{figure}[t]
	\centering
	\input{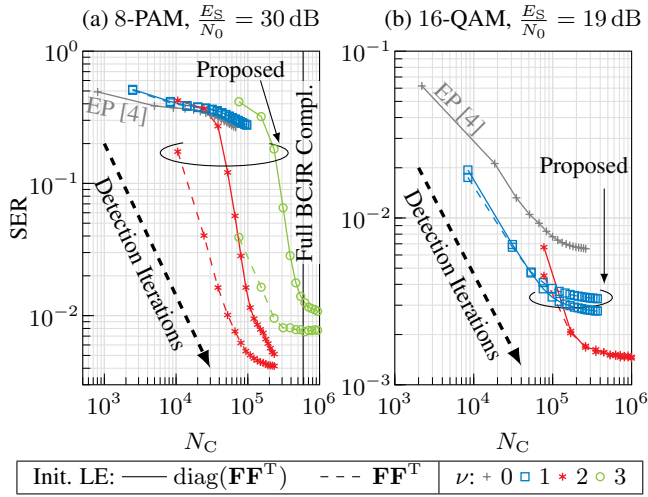}
	\vspace{-0.5em}
	\caption{\small \ac{SER} vs. complexity number $N_\mathrm{C}$, which is the weighted sum of summations, multiplications and Jacobian logarithms per symbol for the (a) Proakis-C and the (b) DICHASUS channel. Each data point resembles an iteration. The line type indicates the covariance matrix used in the \ac{LE} in iteration $\ell=0$. }
	\label{fig:complexity} 
	\vspace{-1.0em}
\end{figure}

For the numerical evaluation of the proposed scheme, we consider two channel models:
First, the well-known Proakis-C channel model \cite{proakis2001digital} with the  static and real-valued impulse response ${\hv=[0.227, 0.460, 0.688, 0.460, 0.227]^\textrm{T}}$ and memory ${L=4}$.
This \ac{CIR} represents the case of strong \ac{ISI} and can be seen as a worst-case \ac{ISI} scenario.
The second channel model is based on the \emph{dichasus-005x} dataset, where $26161$ \acp{CIR} were measured in an indoor non-line-of-sight scenario with the \ac{DICHASUS} \cite{dataset-dichasus-005x}. 
The carrier frequency is \qty{1.272}{GHz} and the bandwidth is \qty{50}{MHz}.
While the \acp{CIR} in the dataset are structured in terms of spatial, frequency, and temporal dimensions, we scramble the \acp{CIR} to evaluate over the average behavior of a single \ac{CIR}.
Further, all taps with an average gain of less than \qty{-20}{dB} are omitted.
This results in a complex-valued channel model with memory $L=8$.
For all evaluations, ${N=512}$ symbols are transmitted.
In our \ac{EP} implementation, variances are clipped at \qty{1e-7}{}, negative variance updates are ignored.

Fig.~\ref{fig:perfromance}(a) displays the performance in terms of \ac{SMI} with an $8$-\ac{PAM} over the Proakis-C channel model.
For each memory $\nu=0,1,2,3$ of the transformed system model, three curves are shown.
The dotted curve corresponds to the initialization of \ac{EP} without iterating ($N_\mathrm{It}=0$), the dashed curve shows the results for a practical choice of parameters ($N_\mathrm{It}=4$, $\beta=0.4$), and the solid curve is the maximum \ac{SMI} optimized over $\beta\in[0.05,0.1,0.2,0.4,0.6]$ for $N_\mathrm{It}=16$ iterations.
We observe that $\nu=1$ does not improve over $\nu=0$, which can be explained by the intuition that $\Fm$ should be ``close'' to $\Hm$ \cite{rusek12optimalshortening}, and two taps cannot well-approximate a symmetric five tap \ac{CIR} like Proakis-C.
Consequently for ${\nu=2}$, we observe a \qty{5}{dB} gain to the original \ac{EP} detector at ${I_\mathrm{SMI}=2.0}$ bits per channel use.
Increasing the memory to $\nu=3$ yields an additional gain of \qty{1}{dB}, reducing the gap to the optimal performance of the \ac{BCJR} detector to \qty{2}{dB}.
Finally, we see that the difference between the practical parameters in dashed and the best possible performance in solid is small.
Analogously, Fig.~\ref{fig:perfromance}(b) shows the \ac{SER} for the \ac{DICHASUS} dataset with $16$-QAM modulation.
At a \ac{SER} of \qty{2e-3}{}, the proposed detector
demonstrates a gain of more than \qty{2}{dB} compared  to the \ac{EP} detector ($\nu=0$) and a gain of \qty{0.7}{dB} comparing the proposed detector with $\nu=2$ and $\nu=1$.
Evaluating a larger $\nu$ or the \ac{BCJR} detector is infeasible.

Fig.~\ref{fig:complexity} shows the \ac{SER} over the complexity number $N_\mathrm{C}$ per symbol, similar to \cite{schmid22neuralBP}, for the cases as in Fig.~\ref{fig:perfromance} at a fixed \ac{SNR}.
Note that $N_\mathrm{C}$ does not resemble the complexity of an actual implementation but serves as an indicator of the order of complexity.
The indicator $N_\mathrm{C}$ is the weighted sum of additions, multiplications, and Jacobian logarithms with weights $1$, $2$, and $2$, respectively.
Furthermore, only online operations during inference are counted in the \ac{LE} and the \ac{NLE}, and edge effects are omitted.
The number of operations in the \ac{LE} is counted as an adaptive filter-type implementation \cite{santos_turbo_2018} with window size $w=3L+1$, and the inverse is assumed to need $w^3$ operations.
The solid curves use a diagonal covariance matrix in the initial \ac{LE} and the dashed curves use the initialization proposed in Sec.~\ref{sec:convergence} with a full covariance matrix in the \ac{LE}.
For the Proakis-C channel in Fig.~\ref{fig:complexity}(a), initialization significantly impacts convergence, whereas for the \ac{DICHASUS} channel in Fig.~\ref{fig:complexity}(b), it does not.
Overall, we can observe better performance for increased $\nu$ at fixed complexity number.

\section{Conclusion}
We proposed an \ac{EP} detector operating in a transformed signal space and evaluated it in terms of performance and complexity.
The algorithm iterates between a linear filter-based estimator and a reduced-complexity \ac{BCJR} detector. 
In addition, we propose a deliberate mismatch between the initial variance passed to the \ac{LE} and the covariance matrix used in the \ac{LE} that improves the convergence speed.
Overall, the results demonstrate significant gains, especially at high rates, and a favorable trade-off between performance and complexity.

\pagebreak

\bibliographystyle{IEEEbib}
\bibliography{references}

\end{document}